\documentclass{article}
\usepackage{latexsym,amsfonts,amssymb, proof, epsf}
\usepackage[dvips]{graphics}
 
\newcommand{\ket}[1]{|{#1}\rangle} 
 
\begin{document}

\title{The internal logic of Bell's states
}

\author
{
Giulia Battilotti and Paola Zizzi
\\
Dipartimento di Matematica Pura ed Applicata
\\
Universit\`a di Padova
\\
via G. Belzoni n.7, I--35131 Padova, Italy
\\
giulia@math.unipd.it
\\
zizzi@math.unipd.it
}
\date{}
\maketitle

\begin{abstract}
We investigate the internal logic of a quantum computer with two qubits, in the two particular cases of
non-entanglement (separable states) and maximal entanglement (Bell's states). To this aim, we consider an internal 
(reversible) measurement which preserves the probabilities by mirroring the states. 
We then obtain logical judgements for both cases 
of separable and Bell's states.
\end{abstract}

\section{Introduction}
The main aim of our work is to look for the internal logic of quantum computation \cite{NC}, illustrating the point of view
of a hypotetical ''internal observer" who lives inside the black box. Such an observer, introduced in \cite{Z}, can 
perform "internal" (reversible) measurements in the quantum system. The idea is that internal measurements give 
rise to logical
assertions \cite{B}, \cite{BZ}, which are then treated following the reflection principle as in basic logic \cite{SBF}. 
By the reflection 
principle, logical connectives are the result of importing some pre-existing metalinguistic links between assertions
into the formal language. We then obtain adequate connectives, corresponding to the 
{\em physical} links which are present inside the black box. 

In \cite{BZ} whe have considered a toy-model quantum computer with one qubit and we have
obtained an 
interpretation of the {\em superposition}
 of the two basis states in terms of the additive conjunction ''$\&$" (and,
dually, with the additive disjunction ''$\oplus$"). The resulting logic is paraconsistent \cite{DC}, \cite{DCG},
\cite{P}, and symmetric, 
like basic logic. We remind that in a paraconsistent logic, both the non-contradiction and the excluded middle principle
do not hold.
Here, we introduce a model of two qubits. This makes it possible to deal with two different physical links 
occurring between two qubits of the register: 
{\em maximal entanglement} (the two qubits are a Bell pair) and {\em non entanglement} (the two qubits state is separable).

\section{Measurements and Mirrors}\label{misure}
To obtain the judgements for the two qubits model, we extend the definition of the internal 
measurement to the case of two qubits. We remind that, in a Hilbert space,
the internal measurement of one qubit is given by a unitary $2\times 2$ complex matrix \cite{Z}.
In such a model, the judgements are obtained by means of a particular internal measurement, called
"mirror measurement" \cite{BZ}, given by the matrices:
\begin{equation}
\label{mirror1}
M=e^{i\phi }\left(\begin{array}{cc}\alpha & 0\\0 & \alpha ^*\end{array}\right) 
\end{equation}
where $\alpha \alpha^*=|\alpha |^2=1$. We have: 
\begin{equation}\label{eq1}
M(a\ket{0} + b\ket{1})=
e^{i\phi}(\alpha a\ket{0} +  \alpha^*b\ket{1})
\end{equation}
So, our mirrors are ''quasi-identities"; actually, they modify the 
longitude of the qubit in the Bloch sphere, that is, the probability amplitudes:
$$
a\rightarrow a'=e^{i\phi}\alpha a
$$
$$
b\rightarrow b'=e^{i\phi}\alpha^* b
$$
and preserve the ''internal truth" given by the probabilities, since $|a'|^2=|a|^2$ and $|b'|^2=|b|^2$.
For this reason, we have chosen mirrors matrices to witness the internal truth and the consequent logical judgements,
as we shall see in the next section.

We now extend the mirror matrices to $C^4$. 
If $M_1=e^{i\phi_1 }\left(\begin{array}{cc}\alpha & 0\\0 & \alpha ^*\end{array}\right)$ and 
$M_2=e^{i\phi_2 }\left(\begin{array}{cc}\beta & 0\\0 & \beta^*\end{array}\right)$, 
the tensor product $M=M_1\otimes M_2$ given by:
\begin{equation}\label{mirror2}
M=
e^{i(\phi_1+\phi_2)}\left(\begin{array}{cccc}
\alpha\beta & 0 & 0 & 0\\
0 & \alpha\beta^* & 0 & 0\\
0 & 0 & \alpha^*\beta & 0\\
0 & 0 & 0 & \alpha^*\beta^*
\end{array}\right)=
e^{i\phi}\left(\begin{array}{cccc}
\gamma & 0 & 0 & 0\\
0 & \delta & 0 & 0\\
0 & 0 & \delta^* & 0\\
0 & 0 & 0 & \gamma^*
\end{array}\right) 
\end{equation} 
is also a mirror matrix. In fact, the most general state  of $C^4$ in the computational basis is:
\begin{equation}\label{reg}
\ket{\psi}
= a\ket{00} + b\ket{01} + c\ket{10} + d\ket{11} 
\end{equation}
and one has:
\begin{equation}\label{eq2}
M\ket{\psi}= e^{i\phi}(\gamma a\ket{00} + \delta b\ket{01} + 
\delta^*c\ket{10} + \gamma^*d\ket{11})
\end{equation}
and again we have: $ a\rightarrow a'=e^{i\phi}\gamma a...$ and so on, then probabilities are preserved: $|a'|^2=|a|^2$...
and so on.

Note that, if $\ket{\psi}$ is one of the Bell states 
$\ket{\psi_{\pm}}=1/{\sqrt 2}(\ket{00} \pm \ket{11})$, its mirroring is:
$M\ket{\psi_{\pm}}=1/{\sqrt 2}e^{i\phi}(\gamma \ket{00} \pm \gamma^*\ket{11})$.
Similarly, for the Bell's state $\ket{\phi_{\pm}}=1/{\sqrt 2}(\ket{01} \pm \ket{10})$, we have
$M\ket{\phi_{\pm}}=1/{\sqrt 2}e^{i\phi}(\delta \ket{01} \pm \delta^*\ket{10})$.
Then Bell states behave as a single particle in the mirroring, since the result has the same form as (\ref{eq1}). 
This fact will be shown in Sect.4 by sequent calculus.
\section{From Mirrors to Judgements in the Black Box}\label{giudizi}
We recall the line of thought followed for the case of the one qubit  model.
Inside the Black Box, a hypothetical internal observer {\bf P} is equipped with 
mirror-matrices and then can perform reversible measurements. Outside the
Black Box, instead, an external observer {\bf G} can perform standard quantum measurements, represented by
projectors. As explained in \cite{BZ}, {\bf G} has a standard quantum logic \cite{BvN}.

It is well known that performing a standard quantum measurement in the given basis, e.g. 
$\ket{0},\ket{1}$ to a qubit $\ket{q}=a\ket{0} + b\ket{1}$, means to apply one of the two projectors $P_0$ or
$P_1$, breaking the superposition and obtaining one of the two basis states. So the observer
{\bf G} can ''read" the value of the qubit as $\ket{0}$, asserting: ''$\ket{0}$ is true"
or $\ket{1}$, asserting: ''$\ket{1}$ is true". So, let us suppose that a standard quantum measurement
is applied and a result $A\in \{\ket{0, \ket{1}}\} $ is obtained. Then {\bf G} asserts "$A$ is true",
written as:
$$
\vdash A
$$
Conversely, denoting by $A^\perp$ the opposite result, {\bf G} asserts "$A^\perp$ is true",
written as:
$$
\vdash A^\perp
$$
By the no cloning theorem \cite{WZ}, after the measurement, {\bf G} can assert {\em only one} of the two. The same 
does not happen to the internal observer {\bf P}, who applies a mirror to $\ket{q}$. In fact, any mirror is the
sum of the two projectors:
\begin{equation}\label{somma1}
M=e^{i\phi }\left(\begin{array}{cc}\alpha & 0\\0 & \alpha ^*\end{array}\right) =
 e^{i\phi}\alpha P_0+ e^{i\phi}\alpha^* P_1
\end{equation}
so that $M\ket{q}=e^{i\phi}\alpha P_0\ket{q}+ e^{i\phi}\alpha^* P_1\ket{q}$. Hence {\bf P} obtains a 
{\em superposition}
of the two results obtainable by {\bf G}. We write then both the above judgements together:
$$
\vdash A \qquad\qquad \vdash A^\perp
$$
What is a couple of possibilities for {\bf G} is instead a unique fact for {\bf P}!  
By the reflection principle, a connective corresponds
to a link between judgements. As in \cite{SBF}, we make the connective ''$\&$" correspond to the
above couple, putting
\begin{equation}\label{eqdef1}
\vdash A\& A^\perp \equiv \qquad\vdash A \qquad \vdash A^\perp 
\end{equation}
Then
\begin{equation}\label{ax1}
\vdash A\& A^\perp  
\end{equation}
''$A\& A^\perp$ is true" is the judgement put by {\bf P} inside the Black Box, concerning the value of the
qubit $\ket{q}$.

Now, let us consider a two-qubit model, that is a Black Box equipped with a register $\ket{\psi}$ of 
two qubits $\ket{q_1}, \ket{q_2}$. Fixed a basis of $C^4$, e.g. the computational basis 
$\ket{00},\ket{01},\ket{10},\ket{11}$,
the external observer, who performs a standard 
quantum measurement in that basis applies one of the four projectors 
$P_{00}, P_{01}, P_{10}, P_{11}$.
Let us suppose that she finds an answer $A\in \{\ket{0},\ket{1}\}$ for $\ket{q_1}$ and an answer
$B\in \{\ket{0},\ket{1}\}$ for $\ket{q_2}$. Then she has a register of two classical bits, and her 
assertion is: 
\begin{equation}\label{giud2}
\vdash A, B
\end{equation}
where the comma stands for the register link between the two classical bits. We interpret the link by the 
multiplicative connective on the right of the sequent, which is called "par", written as "$\oslash$". "Par" has
the same physical meaning of the tensor product "times", written $\otimes$, which, however, is used in linear logic and
basic logic to interpret the comma on the left of the sequent.
Then, as in  \cite{SBF} we put the equation:
\begin{equation}\label{eqdef2}
\vdash A\oslash B\quad \equiv \quad \vdash A, B
\end{equation}
What are all the possible judgements? If the measurements of the two qubits are independent, four combinations are
possible: 
$$
\vdash A, B\quad \vdash A, B^\perp\quad \vdash A^\perp, B\quad \vdash A^\perp, B^\perp
$$ 
and so four judgements
are obtainable:
\begin{equation}\label{quattrogiud}
\vdash A\oslash B\qquad \vdash A\oslash B^\perp\qquad \vdash A^\perp\oslash B\qquad\vdash A^\perp\oslash B^\perp 
\end{equation}
This is the case of a pair of  unentangled qubits. 
On the contrary, let us consider a Bell pair. In such a case the two measurements are  related and so not all combinations
are possible: if $\vdash A, B$ is a result, then $\vdash A^\perp, B^\perp$ is the only other. Note that {\bf G}
is in general unaware of the link existing between $\ket{q_1}$ and $\ket{q_2}$ inside the Black Box, so that the same register
link is used outside. Then, in the case of entanglement, two judgements are possible outside:
\begin{equation}\label{duegiud}
\vdash A\oslash B \qquad\vdash A^\perp\oslash B^\perp 
\end{equation}
As in the case of one qubit, the external observer can put only one of the possible judgements. Again, the judgement of 
the internal
observer is given by a mirror measurement, and mirrors of $C^4$ are obtainable as a linear combination of the four 
projectors:
\begin{equation}
M =e^{i\phi}(\gamma  P_{00} + \delta P_{01} + \delta^* P_{10} + \gamma^* P_{11})
\end{equation}
So {\bf P}, who applies $M$ to the register $\ket{\psi}$, obtains:
\begin{equation}
M \ket{\psi} 
=e^{i\phi}(\gamma P_{00}\ket{\psi} + \delta P_{01}\ket{\psi} + \delta^* P_{10}\ket{\psi} + \gamma^* P_{11}\ket{\psi})
\end{equation}
that is the superposition of the possible values obtainable outside. If $\ket{\psi}$ is not a maximally entangled
state, every projector gives a result and the judgement of the internal observer is obtained as
a superposition of the four judgements in (\ref{quattrogiud}), that is:

\begin{equation}\label{sovrsep}
\vdash (A\oslash B)\&(A\oslash B^\perp)\&(A^\perp \oslash B)\&(A^\perp \oslash B^\perp )
\end{equation}
 If $\ket{\psi}$ is  a Bell state, for example $\ket{\psi_{\pm}}$,
one has $M \ket{\psi_{\pm}} =
e^{i\phi}( \gamma P_{00}\ket{\psi_{\pm}} + \gamma^* P_{11}\ket{\psi_{\pm}})$. The same holds for the other Bell states. 
The result of the measurement of  a maximally entangled state is the superposition of the two judgements
(\ref{duegiud}), that is:
\begin{equation}\label{sovrent}
\vdash (A\oslash B) \&(A^\perp \oslash B^\perp )
\end{equation}
Now, let us consider the internal measurement without any reference to the external one. We know
that {\bf P} achieves a superposition $A\& A^\perp$ measuring $\ket{q_1}$ and another $B\& B^\perp$
measuring $\ket{q_2}$. The two results are linked by two different register links present in the black box,
that is non entanglement and maximal entanglement.
For non entanglement we write $\asymp$, while, for maximal entanglement $\bowtie $. The mirror measurement gives the 
judgements:
\begin{equation}\label{sep}
\vdash (A\& A^\perp)\asymp (B\& B^\perp)
\end{equation}
and \begin{equation}\label{ent}
\vdash (A\& A^\perp)\bowtie  (B\& B^\perp)
\end{equation}
respectively.

We put the two reflection principles, writing $\oslash _0$ (no correlation) and $\oslash _1$ (maximum
correlation) for the two corresponding binary connectives:
\begin{equation}\label{sepeq}
\vdash (A\& A^\perp)\oslash_0 (B\& B^\perp)\equiv\quad\vdash (A\& A^\perp)\asymp (B\& B^\perp)
\end{equation}
and \begin{equation}\label{enteq}
\vdash (A\& A^\perp)\oslash_1 (B\& B^\perp)\equiv\quad\vdash (A\& A^\perp)\bowtie  (B\& B^\perp)
\end{equation}
So we have two kinds of internal judgements concerning our register of two qubits:
\begin{equation}\label{sepgiud}
\vdash (A\& A^\perp)\oslash_0 (B\& B^\perp)
\end{equation}
and \begin{equation}\label{entgiud}
\vdash (A\& A^\perp)\oslash_1  (B\& B^\perp)
\end{equation}
As for the states that are nor separable neither maximally entangled, we argue that connectives 
$\oslash_x$, $x\in (0,1)$ (where $x$ is a degree of correlation), should be introduced, perhaps leading to a kind of 
fuzzy logic.

\section{Towards a  calculus of judgements}
Of course, the internal judgements (\ref{sepgiud}) and (\ref{entgiud}) must be equivalent to the superposition of the external
ones (\ref{sovrsep}) and (\ref{sovrent}), that is we have to prove the following equivalence:
\begin{equation}\label{sepequiv}
\vdash (A\& A^\perp)\oslash_0  (B\& B^\perp)\Longleftrightarrow \;
\vdash(A\oslash B)\& (A \oslash B^\perp) \& (A^\perp\oslash B) \& (A^\perp\oslash B^\perp)
\end{equation}
in the non entangled case, and the equivalence:
\begin{equation}\label{entequiv}
\vdash (A\& A^\perp)\oslash_1 (B\& B^\perp)\Longleftrightarrow \;
\vdash(A\oslash B)\&(A^\perp\oslash  B^\perp)
\end{equation}
in the maximally entangled case.

This can be achieved disassembling the judgements and then assembling them the other way around, as we show in the following
pair of derivations:
\begin{equation}\label{entderiv}
\infer[\& congr] 
{\infer[\oslash_1 ]{\vdash (A\& A^\perp)\oslash_1 (B\& B^\perp)}{\vdash (A\& A^\perp)\bowtie (B\& B^\perp)}
}
{\infer [\&]{\infer[\oslash]{\vdash A, B}{\vdash A\oslash B}\qquad 
\infer[\oslash]{\vdash A^\perp, B^\perp}{\vdash A^\perp\oslash B^\perp}} 
{\vdash (A\oslash B)\& (A^\perp \oslash B^\perp)}
}
\end{equation}
for the maximally entangled case, and
\begin{equation}\label{sepderiv}
\infer[\& cont] 
{\infer[\oslash_0 ]{\vdash (A\& A^\perp)\oslash_0 (B\& B^\perp)}{\vdash (A\& A^\perp)\asymp (B\& B^\perp)}
}
{\infer [\&]{\infer[\oslash]{\vdash A, B}{\vdash A\oslash B}\qquad 
\infer[\oslash]{\vdash A, B^\perp}{\vdash A\oslash B^\perp}\qquad
\infer[\oslash]{\vdash A^\perp, B}{\vdash A^\perp\oslash B}\qquad
\infer[\oslash]{\vdash A^\perp, B^\perp}{\vdash A^\perp\oslash B^\perp}} 
{\vdash (A\oslash B)\& (A \oslash B^\perp)\& (A^\perp \oslash B)\& (A^\perp \oslash B^\perp)}
}
\end{equation}
for the non entangled case. 

We now explain how one must read the above derivations, which are performed by the internal observer.  
The key point, once again, is the superposition link. We remind that, in basic logic, which is a quantum 
linear logic, the additive connective $\&$ is obtained putting the following definitional 
equation:
\begin{equation}\label{basiceq}
\Gamma \vdash A\& B\quad\equiv\quad\Gamma\vdash A \quad \Gamma\vdash B
\end{equation}
which does not admit any context besides the active formulae $A$ and $B$ (visibility of basic logic). 
On the contrary, the same equation in any non-quantum logic can be written with a context $C$:
\begin{equation}\label{classeq}
\Gamma \vdash A\& B, C\quad\equiv\quad\Gamma\vdash A, C \quad \Gamma\vdash B, C
\end{equation}
But, in this case, one would derive distributivity of the multiplicative connective $\oslash$ with respect to the 
additive
connective $\&$ (e.g. see \cite{SBF}).
Inside the black box, we have (at least!) two kinds of distinct links for registers and hence we can deal with
two different versions of the equation for $\&$:
\begin{equation}\label{entsovreq}
\Gamma \vdash (A\& A^\perp)\bowtie (B\& B^\perp) \quad\equiv\quad\Gamma\vdash A, B \quad \Gamma\vdash A^\perp, B^\perp
\end{equation}
for maximal entanglement, and:
\begin{equation}\label{sepsovreq}
\Gamma \vdash (A\& A^\perp) \asymp (B\& B^\perp)\quad\equiv\quad\Gamma\vdash A, B \quad \Gamma\vdash A^\perp, B
\quad \Gamma\vdash A, B^\perp\quad \Gamma\vdash A^\perp, B^\perp
\end{equation}
for non entanglement.

Note that this last equation could be derived from the classical one! 
Then, of course, a classical logician would derive only the rules for the unentangled case, that correspond to a 
classical
use of the context. On the contrary, in the equation (\ref{entsovreq}), we have an odd use of the context, 
corresponding to entanglement. 
Notice moreover that, due to visibility, 
the basic logic equation (\ref{basiceq}) represents a lower bound for both the equations valid in the black box,
that is (\ref{entsovreq}) and (\ref{sepsovreq}). This means again that the external observer, unaware of the actual
kind of correlations present in the black box, but aware of her unawareness, must, as Scepticism  suggests, 
suspend judgement. Hence, for her judgements, no context at all. 

Let's us go back and explain the above derivations (\ref{entderiv}) and (\ref{sepderiv}). 
As we have seen in basic logic, the definitional equations give rise to formation and implicit reflection rules, each one
corresponding to a direction of the equivalence. Such rules can be found in the above couple of derivations, 
which can be read top-down and bottom-up, providing the required equivalences between the judgements. 
The premise $\Gamma$, used in the definitional equations,
is not present in our judgements yet, because its justification inside the Black Box is under study.
In particular, in (\ref{entderiv}) we have the new rule
''$\&congr$", which follows from the definitional equation (\ref{entsovreq}), where ''$congr$" is for ''congruence", 
as equation (\ref{entsovreq}) resembles a congruence rule. It shows in logical terms that entanglement is a
particular form of superposition. In (\ref{sepderiv}) the rule ''$\&cont$" (for "context"), coming from the definitional
equation (\ref{sepsovreq}), that is equivalent to the classical equation (\ref{classeq}), is a form of a classical 
$\&$-rule of sequent calculus.

In our opinion, the derivations, despite their simplicity, are already quite informative for a logical calculus 
which aims
to grasp the efficiency of quantum computation. In fact, they show how the ''quantum parallelism", in the entangled 
case, 
can be obtained by only one half of the derivation branches! This is achieved thanks to  the rule
''$\&congr$". As for the non entangled case, the rule ''$\&cont$" relizes a classical parallelism 
(superposition without entanglement). 

Note that derivation (\ref{entderiv}) has the same form of the following one, that shows how the judgements 
$\vdash A$ and $\vdash A^\perp$ can be assembled and disassembled in the one qubit case (cf. \cite{BZ}):
\begin{equation}
\infer
{\vdash A\& A^\perp}
{\infer{\vdash A \quad \vdash A^\perp}{\vdash A\& A^\perp}}
\end{equation}

In this sense we think that a Bell's state seen from inside the quantum computer can be assimilated to a single
particle. Notice that here ''inside" means that the quantum computer is embedded in a non-commutative geometry 
background \cite{Z}. In other words a 2-qubits register is a fuzzy sphere with four elementary cells each one 
encoding a two-qubits string $\ket{00}, \ket{01}, \ket{10}, \ket{11}$ (see fig. (1)). In the maximally entangled 
case, the fuzzy
sphere has two cells, each one with doubled surface area, and encoding the two-bits string $\ket{00}$ and $\ket{11}$
See fig. (2). 
The latter situation resembles the case of one qubit, where the fuzzy sphere has two elementary cells (each one 
encoding the one bit string $\ket{0}$ and $\ket{1}$ (see fig. (3)). 

The fact that a Bell's state (as seen from inside a quantum computer) ''pretends" to be a single particle, while we 
think it is not, is at the origin of
all paradoxes related to entanglement. For example, ''non-locality" is just a problem of the external observer,
who lives in a local space-time. Instead, the Bell's state, as seen from inside a quantum computer,
lives in a non-local space, which is the fuzzy sphere.
Moreover, as far as causality is concerned, let us consider the cut rule:
$$
\frac{A\vdash B \quad B\vdash C}{A\vdash C}\,cut
$$
which is a causal relation. But the cut rule is not admissible inside the Black Box (as we have seen in \cite{BZ}), 
since it corresponds
to a projective measurement performed in the external world. Thus we argue that the usual meaning of causality is absent 
in the case of a quantum
computer on a fuzzy sphere (internal logic). This is the reason why an external observer, who lives in a causal world,
sees as a paradox the non-causal behaviour of a Bell's state. 

Notice that the model of a quantum computer in a non-commutative geometry can be identified with a model of Computational
Loop Quantum Gravity (CLQG) \cite{Z2}. For a review on Loop Quantum Gravity (LQG) see for example \cite{R}. This is
equivalent to consider a quantum computer at the Planck scale. Causality at the Planck scale is a very controversial
issue, and some authors, like Sorkin and collaborators \cite{S} are inclined to believe in a sort of micro-causality
that they discuss in terms of causal sets. However, the fact is that the light cone at the Plank scale might be
''smeared" by the very strong quantum fluctuations of the metric field (the ''quantum foam" \cite{W}) and this would 
indicate 
that causality is lost 
at that scale. At the light of our logical result, i.e., that the cut rule is not admissible inside the Black Box, we
are now lead to argue that causality is absent at the fundamental scale.

\bigbreak
{\bf Acknowledgements:} 
Work supported by the research project "Logical Tools for Quantum Information Theory", 
Department of Pure and Applied Mathematics, University of  Padova.

\begin{figure}{{\large
{\bf
Fig. 1}}
\medbreak
{\large 
Two unentangled qubits.}}
\label{figura1}
\begin{center}
\includegraphics{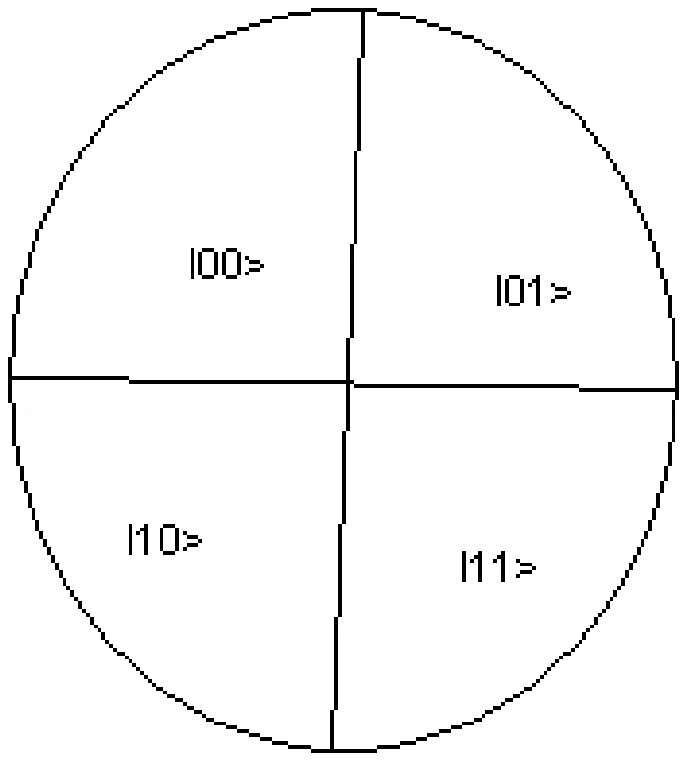}
\end{center}
\end{figure}
\newpage
\begin{figure}{{\large
{\bf
Fig. 2}}}
\medbreak
{\large  A Bell's state}
\label{figura2}
\begin{center}
\includegraphics{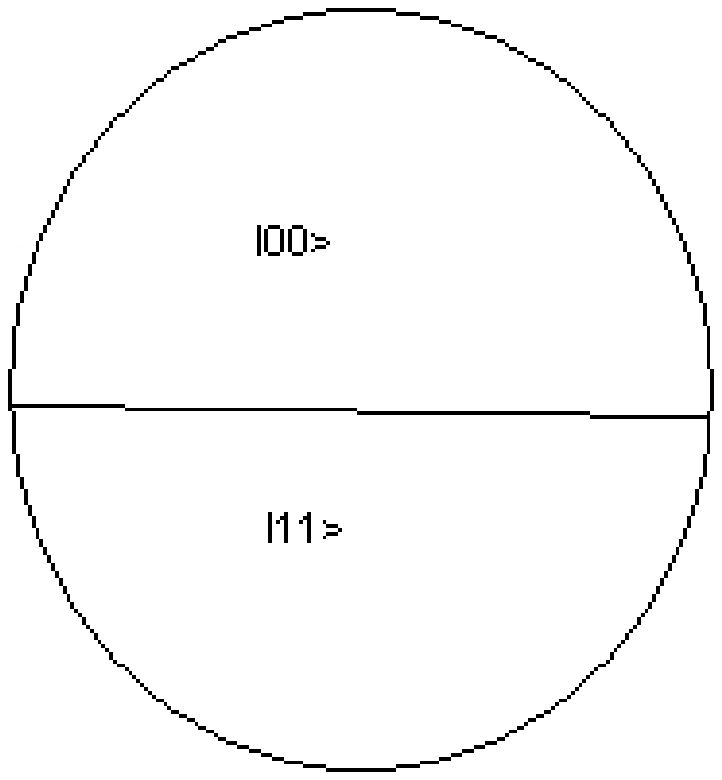}
\end{center}
\end{figure}
\newpage
\begin{figure}{{\large
{\bf
Fig. 3}}}
\medbreak
{\large  One qubit}
\label{figura3}
\begin{center}
\includegraphics{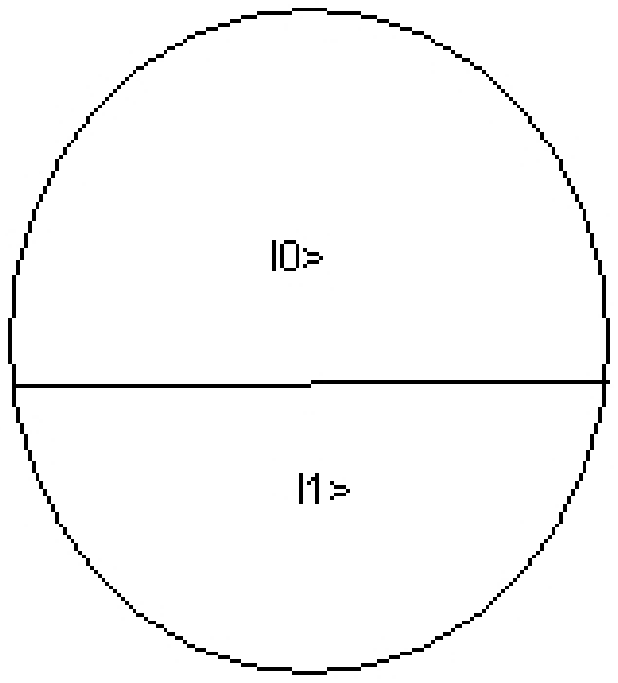}
\end{center}
\end{figure}
\end{document}